\documentclass[aps,footinbib,prl,superscriptaddress,reprint]{revtex4-1}
\usepackage{graphicx}
\usepackage{amssymb}
\usepackage{amsmath}
\usepackage{slashed}
\usepackage{lipsum}
\usepackage[dvipsnames]{xcolor}
\usepackage{xcolor}   
\usepackage{xspace} 
\usepackage{bm}
\usepackage{xfrac}
\usepackage{lineno}
\usepackage[normalem]{ulem}
\usepackage[toc,page]{appendix}
\usepackage{nicefrac}
\usepackage{relsize}
\usepackage{ulem}
\usepackage{adjustbox}
\usepackage{bm}

\usepackage[T1]{fontenc}      
\usepackage{ae} 

\usepackage{grffile}

\RequirePackage{fix-cm} 
\DeclareMathSizes{9.5}{9.5}{7.0}{5}

\usepackage{ytableau} 
\usepackage{slashed} 
\usepackage[enableskew]{youngtab} 
\usepackage{braket}
\usepackage{harpoon}
\usepackage{mathtools}
\usepackage[
	colorlinks=true, 
   filecolor=black,      
   urlcolor=black,
   linkcolor=red,
   citecolor=red]{hyperref}
   \usepackage{capt-of}
\usepackage[capitalize]{cleveref}

\begin{document}
\title{Gauge/gravity dynamics for composite Higgs models and the top mass}
\author{Johanna Erdmenger}
\email{erdmenger@physik.uni-wuerzburg.de}
\affiliation{Institute for Theoretical Physics and Astrophysics, Julius-Maximilians-Universit{\"a}t W{\"u}rzburg,\\
$97074$ W{\"u}rzburg, Germany\\
}
\author{Nick Evans}
\email{n.j.evans@soton.ac.uk}
\affiliation{School of Physics \& Astronomy and STAG Research Centre, University of Southampton, \\
Highfield, Southampton  SO$17$ $1$BJ, UK.\\
}
\author{Werner Porod}
\email{porod@physik.uni-wuerzburg.de}
\affiliation{Institute for Theoretical Physics and Astrophysics, Julius-Maximilians-Universit{\"a}t W{\"u}rzburg,\\
$97074$ W{\"u}rzburg, Germany\\
}
\author{Konstantinos S.~Rigatos}
\email{k.c.rigatos@soton.ac.uk}
\affiliation{School of Physics \& Astronomy and STAG Research Centre,  University of Southampton, \\
Highfield, Southampton  SO$17$ $1$BJ, UK.\\
}
\begin{abstract} 
\noindent We provide gauge/gravity dual descriptions of  explicit realizations of the strong
coupling sector of composite Higgs models using insights from
non-conformal examples of  the AdS/CFT  correspondence.  We calculate
particle masses and pion decay constant for proposed Sp(4) and SU(4) gauge theories, where there is the best lattice data for comparison. Our results compare favourably to
lattice studies and go beyond those due to a greater flexibility in
choosing the fermion content. That content changes the running
dynamics and its choice can lead to sizable changes in the bound state masses.
We describe top partners by a dual fermionic field in the bulk.
Including suitable higher dimension operators  can ensure
a top mass consistent with the Standard Model.
\end{abstract}
\maketitle

Extensions of the AdS/CFT
correspondence \cite{Maldacena:1997re,Witten:1998qj,Gubser:1998bc} to
less symmetric gauge/gravity duals have proved powerful in describing
strongly coupled gauge theories. Adding probe branes allows 
quarks in the fundamental representation of the gauge group to be introduced 
\cite{Karch:2002sh} and to  study the related meson operators \cite{Kruczenski:2003be,Erdmenger:2007cm}. These methods were successfully used to obtain gravity duals of chiral symmetry breaking ($\chi SB$) and pseudo-Goldstone bosons in confining non-Abelian gauge theories \cite{Babington:2003vm,Kruczenski:2003uq}. A natural extension is to apply this approach to other strongly coupled theories in the context of particle physics. Our attention here is on composite Higgs models (CHM) of Beyond the Standard Model physics, as reviewed in  \cite{Cacciapaglia:2020kgq,Panico:2015jxa}. 

The key element of CHM is a strongly coupled gauge theory,  like QCD, causing $\chi SB$ in the fermion sector and  generating four or more Nambu-Goldstone bosons, or ``pions'' \cite{ArkaniHamed:2002qy}. By weakly gauging the global chiral symmetries, four then pseudo-Nambu Goldstone bosons (pNGBs)  can be placed in a doublet of $SU(2)_L$ to become the complex Higgs field. 
The composite nature of the Higgs removes the huge level of fine tuning in the Standard Model (SM) hierarchy problem. This strong dynamics would occur at
a scale of 1-5 TeV, the expected scale for bound states. The LHC has started and, in future runs, will continue to search for such states. 

There is some lattice gauge theory work on CHMs
\cite{Bennett:2019cxd,Bennett:2019jzz,Ayyar:2018zuk,Ayyar:2017qdf,Ayyar:2018glg}. It
is limited though by the cost of numerics and the inability to
unquench fields (i.e.~to include the effect of flavour modes on the gauge
fields) and to match the models' precise fermion content. Here we use
non-conformal gauge/gravity models that explicitly include the gauge
theories' dynamics through the running of the anomalous dimension
$\gamma$ of the fermion or `quark'  mass in the CHM. Our models are inspired by top-down models involving probe D-branes embedded into ten-dimensional supergravity - we  retain as much of the DBI structure as we can in our models to correctly capture the true AdS/CFT dynamics. However, we combine this with a  phenomenological approach
and insert sensible guesses for the running of $\gamma$ based
on perturbation theory. We
then predict some of the mesonic and baryonic spectrum of the
theory. These top-down
inspired holographic models describe the running dynamics of the specific CHM  rather than the more generic previous analyses, e.g.~the Randall-Sundrum \cite{Randall:1999ee}
approach of   \cite{Contino:2003ve,Agashe:2004rs}.

We consider two models: an $Sp(4)$ theory with 4 fundamental and 6 sextet Weyl quarks \cite{Barnard:2013zea}; and an $SU(4)$ theory with five sextet Weyl and 3 fundamental Dirac quarks \cite{Ferretti:2014qta,Ferretti:2016upr}.
Both models can
incorporate a SM Higgs amongst their pNGBs.
 We present results for the
mass spectrum of bound states analogous to mesons in these theories
and compare to relevant lattice results: for the Sp(4) theory, lattice
results were obtained for the quenched theory; for the SU(4) theory,
they have a slightly different fermion content with an even number of multiplets. 
The pattern of masses is well reproduced by the holographic model,
although precise values can differ by up to 20\%.  This gives us
confidence to trust our predictions for how the masses change as we
unquench, including the effect of flavours on the
running coupling, and move to the true fermion content of the
models. In particular, unquenching tends to separate the scales of the
sextet and fundamental matter mesons. Slowing the running also reduces the scalar meson masses \cite{Evans:2013vca}. 

Finally, we address the generation of the top quark's large mass in CHMs, while also keeping flavour changing neutral currents under
control. This can be achieved by the mechanism of partial
compositeness \cite{Kaplan:1991dc} to generate the top-Higgs Yukawa coupling with higher dimension operators (HDOs) from a flavour scale above the strong dynamics scale.   
These CHM use a mechanism where the left
$t_L$ and right handed $t_R$ top quarks mix with baryon-like `top
partner'  spin 1/2 states $T_L, T_R$ in the gauge theory with the same quantum numbers 
\cite{Kaplan:1991dc}.  The top partners are involved in the strong dynamics and so
have an order one Yukawa coupling to the Higgs. HDOs then mix the top and top-partner fields to
generate the top Yukawa coupling. To achieve the large top mass though requires top partner masses of the order of 800 GeV, which is below the natural baryon mass scale and not
consistent with current LHC data
\cite{Aaboud:2018pii,Sirunyan:2019sza}. \footnote{However, these bounds can be lower if non-standard decays are important \cite{Cacciapaglia:2019zmj}.}

We address
this issue by first including  a spinor of appropriate AdS mass into
the holographic model, dual to the top partner operator in the field theory. This allows us to calculate the top partners' mass.   A novel technical ingredient in this respect is the inclusion of spinor fields into a non-supersymmetric  bulk theory, for which we adapt previous results for supersymmetric probe branes \cite{Abt:2019tas}. We also add appropriate strongly coupled HDOs to the holographic model using
Witten's double trace prescription  \cite{Witten:2001ua} . The particular
HDO we pick reduces to a shift in the top partners' mass at low
energies. We show that the top Yukawa coupling can be made of order one
by lowering the top partners' mass to roughly half the CHM's vector
meson mass. This is plausibly reconcilable with experimental constraints.
\vspace{-0.2cm}

\section{Dynamic AdS/YM } \label{sec: QCD}
Our holographic model \cite{Alho:2013dka} is based on the Dirac-Born-Infeld (DBI) action of a top-down model with a D7-brane embedded in a (deformed) AdS$_5$ geometry. The deformation is expanded to quadratic order in the embedding function $X$ (see 
\cite{Alvares:2012kr,Erdmenger:2014fxa}). We add an axial gauge
field as in AdS/QCD models \cite{Erlich:2005qh,DaRold:2005mxj}
and a spinor. The model describes either a single quark in the
background of the gauge fields and other quarks; or, by placing the
fields in the adjoint of flavour and tracing over the action, multiple mass-degenerate quarks. The vacuum $X$ function will cause a global symmetry breaking pattern $G \rightarrow H$.

The model has a dimension one field  for each gauge invariant operator
of dimension three in the  field theory. $X= L e^{i\pi}$ is dual to the complex quark bilinear or any other suitable bilinear operator, depending on the group and representation considered. This field's fluctuations are dual to the analog of the scalar and pseudo-scalar $\sigma$ and $\pi$ mesons of the theory.  $A^\mu_L$ and $A^\mu_R$ are dual to the analogs of the vector and axial vector mesons $V$ and $A$. 

The gravity action of Dynamic AdS/YM is, including also a
spinor field \footnote{$\slashed{D}_{\text{AAdS}}$ is the Dirac operator evaluated
  on  \cref{eq: bckgrnd} }, 
\begin{equation}
\begin{aligned} \label{eq: general_action}
&S= \int d^5 x  \rho^3 \left( \frac{1}{r^2} |D X|^2 + \frac{\Delta m^2}{\rho^2} |X|^2 \right. \\ 
&\left. + \frac{1}{2 g^2_5} \left( F_{L,MN}F_{L}^{MN} + (L \leftrightarrow R) ~ \right) \right. \\ 
&\left. + \bar{\Psi} \left( \slashed{D}_{\text{AAdS}} - m \right) \Psi  \right)\, . 
\end{aligned}
\end{equation}
The five-dimensional coupling is obtained by matching to the UV vector-vector correlator 
$
g_5^2 = \frac{24 \pi^2}{d(R)~N_{f}(R)} \label{eq:grouptheory} \, 
$ \cite{Erlich:2005qh}.
$d(R)$ is the dimension of the quark's representation and N$_f$(R) is the number of Weyl flavours in that representation.
The model lives in a five-dimensional asymptotically AdS (AAdS) spacetime
\begin{align} \label{eq: bckgrnd}
ds^2 = r^2 dx^2_{(1,3)} + \frac{d \rho^2}{r^2} \, ,
\end{align} 
with $r^2 = \rho^2 +L^2$ the holographic radial direction,
corresponding to the energy scale, and $dx^2_{(1,3)}$  a
four-dimensional Minkowski spacetime. The $\rho$ and $L$ factors in
the action and metric are implemented directly from the top-down
analysis of the D3/probe-D7 system. They ensure an appropriate UV
behaviour. In the IR, the fluctuations know about any $\chi SB$ through $L \neq 0$. 

The dynamics of a particular gauge theory, including quark contributions to any running coupling, is included through $\Delta m^2$ in \cref{eq: general_action}. To find the theory's vacuum, with a non-zero chiral condensate,  we set all fields to zero except for $L(\rho)$. For $\Delta m^2$ a constant, the equation of motion obtained from \cref{eq: general_action} is
\begin{align}
\partial_{\rho} (\rho^3 \partial_{\rho} L) - \rho ~ \Delta m^2 L = 0 \label{eq: vacuum qcd} \, .
\end{align}
The solution takes the form $L(\rho) = m \rho^{-\gamma} + c\rho^{\gamma-2}$, with 
$\Delta m^2 = \gamma (\gamma -2)$  in units of the inverse AdS radius
squared. Here $\gamma$ is the anomalous dimension of the quark
mass. The Breitenlohner-Freedman bound \cite{Breitenlohner:1982jf}, below which an instability to $\chi SB$ occurs, is given by $\Delta m^2=-1$. 

We impose a particular gauge theory's dynamics by using its running
$\gamma$ to determine $\Delta m^2$ - as usual in holography,
\mbox{$M^2=\Delta(\Delta-4)$}, with $\Delta$ the operator scaling dimension.  For $\gamma < 1$, we find $\Delta m^2 = - 2 \gamma$, thus a theory triggers $\chi SB$ if $\gamma$ passes through $1/2$. Since the true running of $\gamma$ is not known non-perturbatively, we extend the perturbative results as a function of renormalization group scale to the non-perturbative regime. To find the running of $\gamma$, we use the perturbative results. To provide some understanding of error range in the holographic model, we compute with  both the
one-loop result $\gamma_{1l} = \frac{3~C_2(R)}{2 \pi}~\alpha$ 
and the two-loop result \\ \mbox{$\gamma_{2l} = \frac{\alpha}{2 \pi^2} ({3 \over 2} C_2(R)^2 + {97 \over 6} C_2(R) C_2(G)-{10 N_f\over 3} C_2(R)T(R))$}) with a running $\alpha$ \footnote{For all the group theory factors we have used the Mathematica application LieART \cite{Feger:2012bs}}. 
Note that we decouple fields from the running as they go on mass shell. We average the results and use half the range as the error.

We numerically solve (\ref{eq: vacuum qcd}) with our
ansatz for $\Delta m^2$, using the IR boundary conditions
\mbox{$L(\rho=\rho_{IR}) = \rho_{\text{IR}}$}, $\partial_{\rho}
L(\rho=\rho_{IR}) = 0$. They are imported from the D3/D7  system but imposed at the scale where the quarks go on mass shell. The spectrum is determined by considering fluctuations in all
fields of \eqref{eq: general_action} about the vacuum \cite{Alho:2013dka} - see the supplementary materials for details.

We will use this model to explore two gauge theories that have been
proposed to underlie CHM - we pick the two with the most extensive lattice data. We denote models by their gauge group and the number of Weyl flavours in the fundamental ($F$), and two-index antisymmetric ($A_2$) representations. \medskip

\noindent {\bf Sp(4) gauge theory  - $\bm{Sp(4) ~4F,6A_2}$:}
This CHM, proposed in \cite{Barnard:2013zea}, has top
partners. Lattice studies related to this model were performed in \cite{Bennett:2019cxd,Bennett:2019jzz}.
The fundamental of $Sp(2N)$ is pseudo-real, hence the $SU(2)_L\times SU(2)_R$ symmetry of a model with four Weyl
quarks in the

\begin{table}[!htb]
 \begin{tabular}{c|cccccc}  
                                    & AdS/$Sp(4)$  		& AdS/$Sp(4)$ 	& lattice \cite{Bennett:2019cxd}  	& lattice \cite{Bennett:2019jzz} 	 \\ 
     								& unquench			& quench 		& quench  							& unquench 		  \\    \hline
    $f_{\pi A_2}$  				  	& 0.118 (02) 		& 0.104 (02)	& 0.1453(12) 						&  											\\
    $f_{\pi F}$  					& 0.068 (02) 		& 0.074 (02)	& 0.1079(52) 						& 0.1018(83) 								\\
    $M_{V A_2}$ 					& 1* 				& 1*			& 1.000(32) 						& 												\\
    $M_{V F}$ 					   	& 0.783 (31) 		& 0.92 (04)	& 0.83(19)							& 0.83(27) 								\\
    $M_{A A_2}$ 				 	& 1.35 (01) 		& 1.29 (02)		& 1.75 (13)							& 												\\
    $M_{A F}$ 					 	& 1.16 (03) 		& 1.32 (04)	 	&  1.32(18) 						& 1.34(14) 								\\
    $M_{S A_2}$ 				 	& 0.37 (01) 		& 0.98 (16)	& 1.65(15)	$^\dagger$						& 												\\
    $M_{S F}$ 					 	& 0.77 (13) 		& 1.10 (15)		& 1.52 (11)$^\dagger$							& 1.40(19)$^\dagger$								\\
    $M_{T A_2}$ 					& 1.85 (01)			& 1.85 (05)	&									& 				\\
    $M_{T F}$ 						& 1.46 (07) 		& 1.71 (08)		& 									&					\\
 \end{tabular}   
  
  \caption{ \label{Sp4table} AdS/$Sp(4) ~4F,6A_2$. Ground state masses
    for the vector, axial-vector, and scalar mesons for the two
    representations, $F$ and $A_2$, from  holography
    (errors show the range in using $\gamma_{1l}$ and
      $\gamma_{2l}$)  and the lattice. Also  the pion decay constant
    and two estimates of the top partner baryon mass. We have rescaled
    the lattice data to the $M_{VA_2}$ mass in the quenched case
    \cite{footnote} and for the unquenched case, where no $A_2$ states
    are included, rescaled so $M_{VF}$ matches the quenched case for
    comparison. $\dagger$ Note the lattice scalar
      results are for the $a_0$, not for the $\sigma$ we compute in the holographic model.   \vspace{-0.3cm}
  }
  
\end{table}  

\noindent fundamental representation is enhanced to an $SU(4)$  flavour symmetry \cite{Lewis:2011zb}. The condensation pattern is the
same as 
 in QCD, 
and breaks the $G=SU(4)$ flavour symmetry to $H=Sp(4)$ with 5 pNGBs.

The top partners are introduced by including three additional Dirac fermions in the $A_2$ representation of the gauge group for $N>1$. It is natural to concentrate on the minimal, $Sp(4)$, gauge group case.  The top partners are $F A_2 F$ bound states. From the point of view of the $Sp(4)$ dynamics, there is an ${G=SU(6)}$ symmetry on the 6 $A_2$ Weyl fermions that are in a real representation. The $A_2$ condensate breaks this symmetry to $H=SO(6)$.

The $A_2$ fermions condense ahead of the fundamental fields, since the critical value for $\alpha$ where $\gamma=1/2$  is smaller (at the level of the approximations we use the critical couplings are  \mbox{$\alpha_c^{A_2} =  {\pi \over 6} = 0.53$},  \mbox{$\alpha_c^{F} = {4 \pi \over 15} = 0.84$}).

We perform the AdS/YM analysis for the two fermion sectors separately, although they are linked since both flavours contribute to the running of $\gamma$ down to their IR mass scale. We find the $L(\rho)$ functions for the $A_2$ and $F$ sectors. We fluctuate around each embedding separately to find the spectrum (neglecting any mixing). We present the holographically computed spectrum in \cref{Sp4table}. The errors show the range
in using $\gamma_{1l}$ or $\gamma_{2l}$ and are mostly small. The scalar masses are  dependent on the strength of the
running and so have the largest errors.

Lattice data exists for the quenched theory
\cite{Bennett:2019cxd}. To compare, we provide holographic results
for the quenched case - here no fermions contribute to the
running. The V meson masses fit the lattice data well. For the
A mesons, the holographic model predicts close to degeneracy between
the F and $A_2$ sectors, while the lattice has a wider

\begin{table}[!htb]
\begin{tabular}{c|ccc}  
     						& Lattice $SU(4)$ \cite{Ayyar:2018zuk} 	& AdS/$SU(4)$ 				& AdS/$SU(4)$    \\
     						& $ 4A_2, 2 F, 2 \bar{F} $  			& $ 4A_2, 2 F, 2 \bar{F} $ 	& $ 5A_2, 3 F, 3 \bar{F} $  \\ 
\hline
$f_{\pi A_2}$				& 0.15(4) 								& 0.105 (06)					& 0.111 (01) 		\\
$f_{\pi F}$ 				& 0.11(2) 								& 0.094 (01)					& 0.108 (01)			\\
$M_{V A_2}$ 				& 1.00(4) 								&	1*						& 1* 		\\
$M_{V F}$ 					& 0.93(7) 								&	0.91 (03)					& 0.88 (02) \\
$M_{A A_2}$ 				&   				 					&	1.38 (01)					& 1.33 (01)  \\
$M_{A F}$ 					&  										&	1.35 (02)					& 1.21 (02)		 \\
$M_{S A_2}$ 				&  		  								&	0.746 (13)					& 0.61 (08) 			   \\
$M_{S F}$  					&  										&	0.829 (19)					& 0.66 (01)				\\
$M_{T A_2}$ 				& 1.4(1) 		  						&	1.85 (01)					& 1.84 	(01)			\\
$M_{T F}$ 					& 1.4(1) 		 						&	1.69 (06)					& 1.63 	(05)			\\
  \end{tabular} 
  
    \caption{\label{tab: results_su4_lattice_two_runnings} SU(4)  theories - the spectrum in two scenarios and lattice data for comparison.  \vspace{-0.4cm}
}
\end{table}  

\noindent  spread (although with large errors on the spread) and a 20\% higher estimate
of the scale.  The holographic pion decay constants are low by
20-30\%. The lattice results are for  the $a_0$
  masses, not for the $\sigma$, but we include these data points for comparison. Overall, the success in
the pattern suffices to make us trust changes as the theory is
unquenched. Particularly, the gap between quantities in the $A_2$ and
$F$ sectors grow by 10-20\% as the running between the condensation
scales slows due to the $A_2$ being integrated out, and the scalar masses fall as the running 
 slows. There are unquenched lattice results
but only for the F sector \cite{Bennett:2019jzz}, so they do not shed light on the mass gap between the sectors or the dependence of the $\sigma$ mass.  

$\left. \right.$ \vspace{-0.2cm}

\noindent{\bf SU(4) gauge theory - SU(4)  $\bm{3F, 3 \bar{F}, 5 A_2}$: } 
We report on a second model \cite{Ferretti:2014qta,Ferretti:2016upr},
for which there has also been related lattice work. The gauge group is
$SU(4)$. There are 5 Weyl fields in the $A_2$ representation. When
these $A_2$ condense they break their $G=SU(5)$ symmetry to $H=SO(5)$ -
the pNGBs include the Higgs. To include top partner baryons, fermions
in the fundamental are added, allowing $F A_2 F$ states. The number of
flavours is fixed to be three Dirac spinors, since we need to be able
to weakly gauge the flavour symmetry to become the SU(3) of QCD. When
these fields condense, the chiral $G=SU(3)_L \times SU(3)_R$ symmetry is
broken to the vector $H=SU(3)$ subgroup. Note that this model is hard to
simulate on the lattice due to the fermion doubling problem, so
instead lattice work \cite{Ayyar:2017qdf,Ayyar:2018zuk} has focused on
the case with just 2 Dirac $A_2$s and 2 fundamentals, which allows 
consideration of the unquenched case.


We display our holographic results in (\ref{tab:
  results_su4_lattice_two_runnings}) with lattice data from
\cite{Ayyar:2018zuk}. The holographic approach and the lattice agree
in the V sector and the pion decay constants lie close to or just
below the lower lattice error bar. We obtain additional masses that
have not been computed on the lattice. Crucially, we move to the $
5A_2, 3 F, 3 \bar{F} $ fermion content by adding in the additional
fields needed for the  CHM of
\cite{Ferretti:2014qta,Ferretti:2016upr}. We see that the extra fermions slow the running between the $A_2$ and $F$ mass scales, reducing the $F$ sector masses by 10\%. The scalar masses again fall (by 20\%) in response to the slower running.
\vspace{-0.4cm}

\begin{figure}[!htb]
 \centering
 \begin{adjustbox}{center}
   \includegraphics[height=45mm,width=0.9\columnwidth]{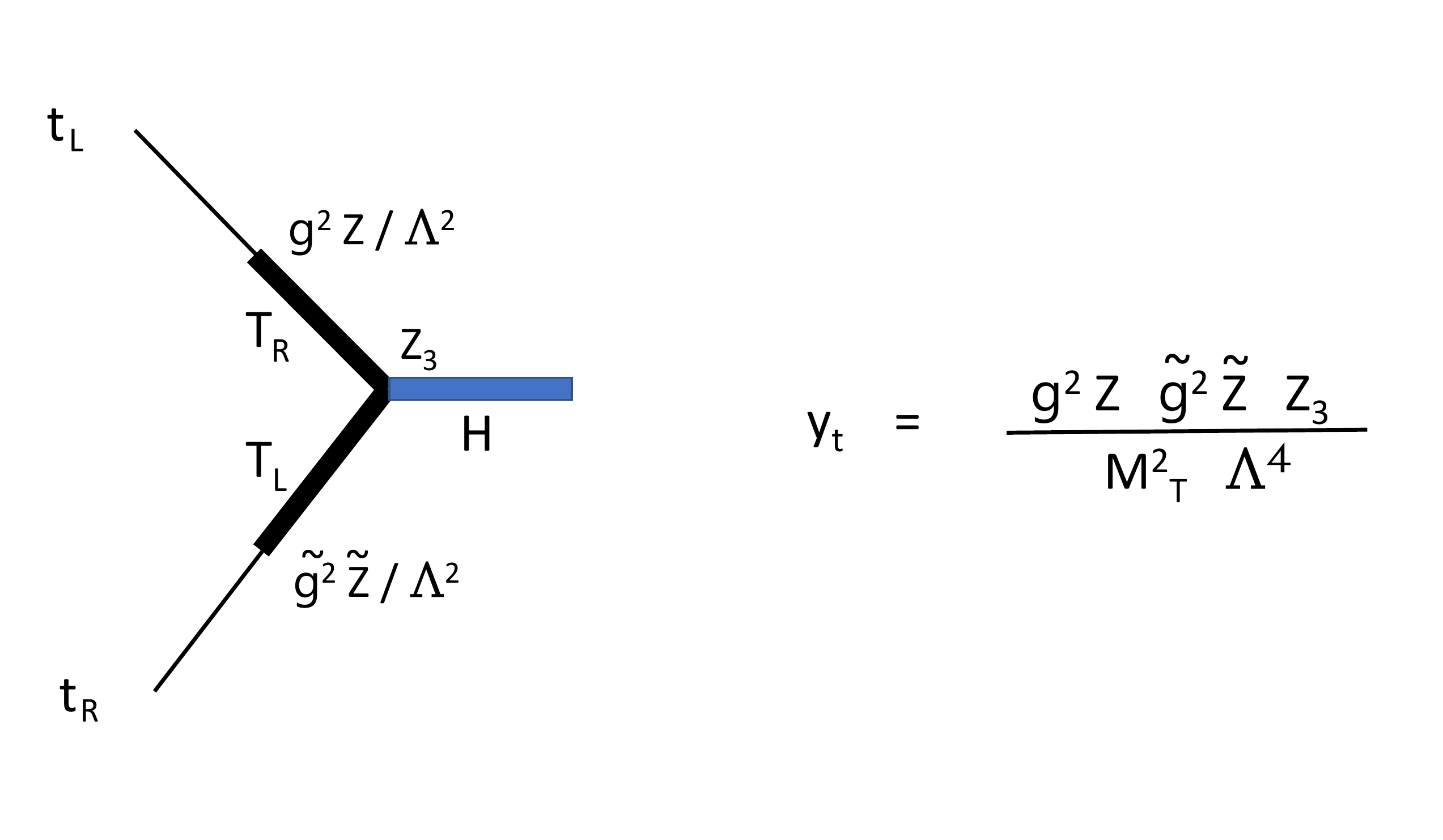}
 \end{adjustbox}
 \vspace*{-10mm}
\caption{The top Yukawa coupling  ($y_t$) vertex diagram. \vspace{-0.2cm}}
 \label{topdiagram}
\end{figure} \vspace{0.2cm}

\section{Top partners} \vspace{-0.2cm}

Top partners in both models we consider are $F A_2 F$ states, spin-$1/2$ baryons of the strong dynamics. We  describe them by a spinor fluctuating in AdS \cite{Kirsch:2006he}, dual to  the baryon operator. From (\ref{eq: general_action}), we derive a second  order wave equation for the spinor, given in \cite{Erdmenger:2020flu}, based  on \cite{Abt:2019tas}. The main features are:
States with $\Delta=9/2$, appropriate for three-quark states, have an AdS mass $m=5/2$. The spinor can be written as eigenstates of the $\gamma_\rho$ projector, with $\gamma_\rho \psi_{\pm} = \pm \psi_{\pm}$. The equation then splits into two copies of the dynamics (for $\psi_+$ and for $\psi_-$)  with explicit relations between the solutions \cite{Abt:2019tas}. One can solve one equation and from the UV behaviour extract the source $\mathcal{J}$ and operator $\mathcal{O}$ values. See the supplementary materials for further details.

This allows the baryon masses' computation for three-fermion states associated with a particular background $L(\rho)$. Here we wish to set  ${\cal O}=F A_2 F$. However, since it is a mixed state  in principle the computation should take into account that the $A_2$ and $F$ states have distinct $L(\rho)$. We simplify this by using one $L(\rho)$, either that for $F$ or $A_2$, which assumes the flavours are degenerate - the two choices give predictions for the baryon masses, $M_{TA_2}$ and $M_{TF}$ at the bottom of \cref{Sp4table,tab: results_su4_lattice_two_runnings}. We expect the mass to lie between these two computations.  There is only lattice data for the SU(4) model for the top partner - our estimate is $25\%$ high, but sets a sensible figure from which to observe changes as we include HDOs.

The top mass itself is generated in CHM by the diagram in
\Cref{topdiagram}.
The $Z$ factors are structure functions that depend on
the strong dynamics.  $g^2/{\Lambda^2}$ and the tilded vertex, which we won't distinguish henceforth, are the dimensionful couplings of the HDOs that mix the top and top partners of the form $\bar{t}_{L/R} {\cal O}$.

On dimensional grounds, 
a sensible holographic  estimate for the $Z_3$ factor is a weighted
integral over the pNGB and baryon wave functions \footnote{
To compute the $Z_3$ factor, we would need a cubic term
in \eqref{eq: general_action}. However, its coupling is not determined in the
bottom-up approach. We use the estimate given in \eqref{eq: top_mass_1a}.},
\begin{equation} \label{eq: top_mass_1a}  
Z_3 \simeq \int d \rho~ \rho^3 { \partial_\rho \pi ~ 
  \psi_B^2 \over ( \rho^2 +L^2)^2} \, .
\end{equation}
A similar contributing term to the Z and $\tilde{Z}$ factors is
\begin{equation} \label{eq: top_mass_2a}
Z \simeq \tilde{Z} \simeq  \int d \rho~
\rho^3 \partial_\rho\psi_B \, .
\end{equation}
$\pi$ and $\psi_B$ are $\rho$ dependent holographic wavefunctions for
the pNGB and baryon, respectively. The $Z$  factor
  was computed on the lattice in \cite{Ayyar:2018glg} - for QCD it is
  expected that $Z\simeq 19 f_\pi^3$ and we find $Z=22.6 f_\pi^3$. For
  the SU(4) lattice variant, $Z \simeq 7.5 f_\pi^3$ is found, and we find $Z= 7.0 f_\pi^3$ showing that our estimate is sensible. 

We have then computed $y_t$ from the full set of factors in \cref{topdiagram} in both CH models. If we set a cut off for the HDOs roughly 6 times the vector meson mass, we find the top Yukawa coupling is only of order 0.01, far below the needed value of 1. This is the standard problem  when trying to generate the top mass - it is suppressed both by the HDO scale $\Lambda$ and top partner mass squared. 

Our new solution to this is to enhance $y_t$ by including  a further
HDO given by
\begin{equation}  \label{Tphdo}
{\cal L}_{\mathrm{HDO}} = {g_T^2 \over \Lambda^5}  |F A_2 F|^2 \, .
\end{equation}
As the operator $F A_2 F$ becomes the top partner, this is directly a shift in its mass. 
To include this holographically, we use Witten's double trace
prescription \cite{Witten:2001ua}, according to which the vacuum
expectation value $\langle
F A_2 F \rangle$ contributes to the source via $\mathcal{J} = {g_T^2
  \over \Lambda^5} \langle F A_2 F \rangle$ once (\ref{Tphdo}) is
turned on. From the asymptotic boundary behavior of the gravity
solutions, we read off $\mathcal{J} $ and $\mathcal{O}$ and then compute $g_T$. From these results we can find the masses of the top partner for a particular $g_T$.

The HDO  (\ref{Tphdo}) can indeed be used to reduce the top partner's mass as was shown in a more formal setting \cite{Abt:2019tas} - for small $g_T$ the effect is linear and small, but after a critical value the effect is much larger. The same conclusion was reached in a simple bottom-up model for the scalar mode as well \cite{Buyukdag:2019ouy}. In the two models above, using either of the baryon mass' estimates, one can reduce the top mass to half the V meson mass (which is likely 0.5TeV or above in CHM) for $7 < g_T <10$.  

Caution is needed in computing the top Yukawa. As the top partner's mass changes, so do the $Z$ factors in (\ref{eq: top_mass_1a}), (\ref{eq: top_mass_2a}). In particular, the HDO in (\ref{Tphdo}) plays a large role since it induces a sizable non-normalizable piece in the UV holographic wave function of the top partner. 
The integrals in the normalization factors for the state, which enter
directly in the expressions for the $Z$ factors, are more dominated by
the UV part of the integral. The overlap between different states can
also change substantially. We therefore plot an example of the full
expression for the Yukawa coupling from \cref{topdiagram} against the
top partner mass (which changes with $g_T$) in \cref{fig:
  spin_1_2_dt_su4_Z}. We see that the top Yukawa grows as the top
partner's mass falls and can become of order 1 as the top partners
mass falls to about half of the vector meson mass. This suggests one might be able to realize a phenomenological viable top partner mass of 1 TeV or so and the required top mass. 

\begin{figure}
\centering
\includegraphics[scale=0.6]{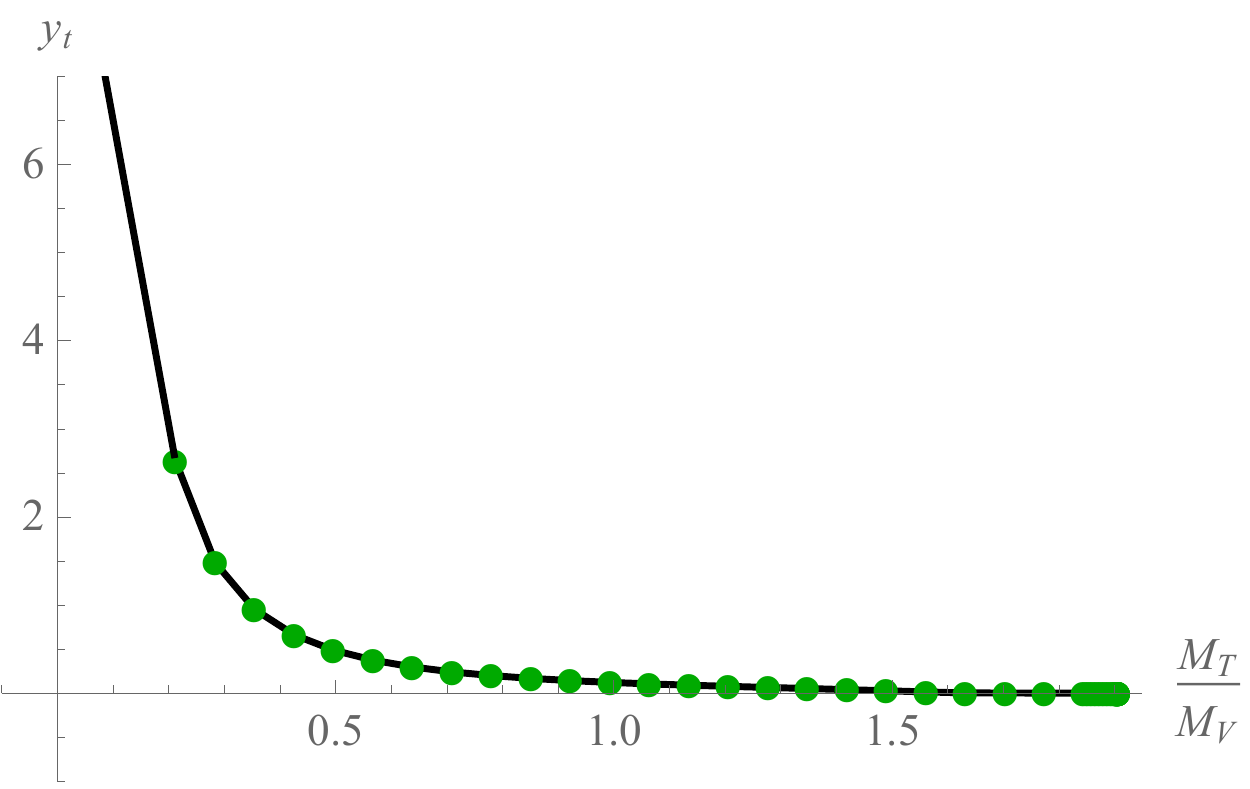} 
\caption{\label{fig: spin_1_2_dt_su4_Z} Yukawa coupling as function of
  the top partner mass in the $A_2$ sector of the SU(4) gauge theory (here $g=\tilde{g}=1$). }
\end{figure}

To conclude, we have shown that gauge/gravity duality methods are a powerful tool for obtaining
sensible estimates in strongly coupled  theories relevant to CHM,
and are thus a resource for model builders.
The AdS/YM theory presented is fast to compute with. The
fermion content  can be changed in a
simple way to obtain further models.  Our results for CHM (and a much wider set of models that can be found in \cite{Erdmenger:2020flu}) suggest how the
spectrum will change as lattice simulations are unquenched to the
correct flavour content and HDOs  introduced.  \vspace{0.1cm}

\noindent {\bf Acknowledgements.}~We thank Raimond Abt, Alexander Broll, Tony Gherghetta and Yang Liu for discussions. NE was supported by the STFC grant ST/P000711/1.  \vspace{-0.4cm}

\section{Supplemental Material}

The meson spectra of the theory can be found by solving the equations
of motion for fluctuations in the fields in Eq. 1 in the main text about the vacuum configuration $X=L_0$, the solution of eq. 3 in the main text. A generic fluctuation is written as $F(\rho) e^{-ik.x}, ~
M^2 = -k^2$ and IR boundary conditions $F(L_{IR})=1, F'(L_{IR})=0$
are used. We solve the fluctuation equations by numerical shooting and seek the values of $M^2$
where the  solution falls to zero in the UV. 

\noindent {\bf Scalar mesons} are obtained from the fluctuation $S$ in   $L= L_0+S$ and the equation of motion is
\begin{equation} \label{eq: eqn of motion_scalar} \begin{array}{c}
\partial_{\rho} (\rho^3 \partial_{\rho} S(\rho)) - \rho (\Delta m^2) S(\rho) - \rho L_{0} S(\rho) \frac{\partial \Delta m^2}{\partial L} |_{L_{0}} \\ + M^2 \frac{\rho^3}{r^{4}} S(\rho) = 0. \end{array}
\end{equation}

\noindent {\bf Vector mesons} are  fluctuations of the gauge fields $V=A_L+A_R$  and satisfy the equation of motion
\begin{align} \label{eq: eqn of motion_vector}
\partial_{\rho} (\rho^3 \partial_{\rho}V(\rho)) + M^2_{V} \frac{\rho^3}{r^{4}} V(\rho) = 0.
\end{align}
To obtain a canonically normalized kinetic term for the vector meson, we must impose
$
\int d\rho~ {\rho^3 \over g_5^2 r^4} V^2 = 1. $

\noindent {\bf Axial mesons} ($A=A_L-A_R$) are obtained from
\begin{equation}  \label{eq: eqn of motion_axial}
\partial_{\rho} (\rho^3 \partial_{\rho} A(\rho)) - g_5^2 \frac{\rho^3 L^2_{0}}{r^2} A(\rho) + \frac{\rho^3 M^2_{A} }{r^{4}} A(\rho) = 0.
\end{equation}

\noindent The {\bf pion decay constant} can be extracted from the axial-vector current two-point function that is related to the decay constant by  $\Pi_{AA} = f_\pi^2$, with 
\begin{equation} f_\pi^2 = \int d \rho {1 \over g_5^2}  \partial_\rho \left[  \rho^3 \partial_\rho K_A(q^2=0)\right] K_A(q^2=0)\,.
\end{equation} 
The axial  source is described as a fluctuation with  non-normalizable UV behavior. In the UV we expect 
$L_0(\rho) \sim 0$ and we  solve
the equations of motion for the scalar obtained from $A^\mu= \epsilon^\mu K_A(\rho) e^{-i q.x}$  which satisfies
\begin{equation}  \label{thing}
\partial_\rho [ \rho^3 \partial_\rho K_A] - {q^2 \over \rho} K_A= 0\,. 
\end{equation}
The solution  is normalized to
\begin{equation} \label{Ks}
K_A = \left( 1 + {q^2 \over 4 \rho^2} \ln (q^2/ \rho^2) \right)
.
\end{equation}
 
\noindent {\bf The baryon}  (mass $M_B$) is described by a 
four-component bulk fermion satisfying 
\begin{equation} \label{fermioneom}
\left( \partial_{\rho}^2 + \mathcal{P}_{1} \partial_{\rho} +
  \frac{M^2_{B}}{r^4} + \mathcal{P}_2 \frac{1}{r^4} -  \frac{m^2}{r^2}
  - \mathcal{P}_3 \frac{m}{r^3} ~ \gamma^{\rho}   \right) \psi = 0 
  \end{equation}
  with
  \begin{equation}
  \begin{array}{l}
\mathcal{P}_1 = \frac{6}{r^2} \left( \rho + L_0~ \partial_{\rho}L_0
  \right) \, , \\
\mathcal{P}_2 = 2 (\rho^2 + L_0^2) L \partial^2_{\rho} L_0 + (\rho^2 +
  3L_0^2) (\partial_{\rho}L_0)^2  \\
\left. \right. \hspace{1.5cm}+ 4 \rho L_0 \partial_{\rho}L_0 + 3 \rho^2 + L_0^2 \, , \\
 \mathcal{P}_3 = \left( \rho + L_0 ~ \partial_{\rho}L_0   \right) \, .
\end{array}
\end{equation}
For states of UV dimension $9/2$, the bulk fermion mass is $m=5/2$. 

The four component spinor can then be written in terms of eigenstates
of $\gamma_\rho$ such that $\psi = \psi_+ \alpha_+ + \psi_- \alpha_-$,
where  $\gamma_\rho \alpha_{\pm} = \pm \alpha_{\pm}$. The equation of motion \cref{fermioneom} then becomes two equations, one for $\psi_+$ and one for $\psi_-$, obtained by replacing $\gamma_\rho$  by $\pm 1$ respectively. The two equations are thus copies of the same dynamics with explicit relations between the solutions. The UV asymptotic form of the solutions are given by 
\begin{align} 
\begin{split}
\psi_{+} &\sim \mathcal{J} \sqrt{\rho} + \mathcal{O} \frac{M_{B}}{6}
{\rho}^{-11/2} \, , \\
\psi_{-} &\sim \mathcal{J} \frac{M_{B}}{4} \frac{1}{\sqrt{\rho}} +
\mathcal{O} {\rho}^{-9/2} \, .
\end{split}
\end{align} 
Here we use the D3/probe D7-brane system [28] as a guide to impose the IR boundary conditions
\begin{align}
\begin{aligned}
\psi_{+}(\rho = L_{IR}) &= 1, &&& \partial_{\rho} \psi_{+}(\rho =
L_{IR}) &= 0 \, ,\\
\psi_{-}(\rho = L_{IR}) &= 0, &&& \partial_{\rho} \psi_{-}(\rho =
L_{IR}) &= \frac{1}{L_{IR}} \, .
\end{aligned}
\end{align}


\bibliographystyle{apsrev4-1} 
\bibliography{lit}

\begin{thebibliography}{45}%
\makeatletter
\providecommand \@ifxundefined [1]{%
 \@ifx{#1\undefined}
}%
\providecommand \@ifnum [1]{%
 \ifnum #1\expandafter \@firstoftwo
 \else \expandafter \@secondoftwo
 \fi
}%
\providecommand \@ifx [1]{%
 \ifx #1\expandafter \@firstoftwo
 \else \expandafter \@secondoftwo
 \fi
}%
\providecommand \natexlab [1]{#1}%
\providecommand \enquote  [1]{``#1''}%
\providecommand \bibnamefont  [1]{#1}%
\providecommand \bibfnamefont [1]{#1}%
\providecommand \citenamefont [1]{#1}%
\providecommand \href@noop [0]{\@secondoftwo}%
\providecommand \href [0]{\begingroup \@sanitize@url \@href}%
\providecommand \@href[1]{\@@startlink{#1}\@@href}%
\providecommand \@@href[1]{\endgroup#1\@@endlink}%
\providecommand \@sanitize@url [0]{\catcode `\\12\catcode `\$12\catcode
  `\&12\catcode `\#12\catcode `\^12\catcode `\_12\catcode `\%12\relax}%
\providecommand \@@startlink[1]{}%
\providecommand \@@endlink[0]{}%
\providecommand \url  [0]{\begingroup\@sanitize@url \@url }%
\providecommand \@url [1]{\endgroup\@href {#1}{\urlprefix }}%
\providecommand \urlprefix  [0]{URL }%
\providecommand \Eprint [0]{\href }%
\providecommand \doibase [0]{http://dx.doi.org/}%
\providecommand \selectlanguage [0]{\@gobble}%
\providecommand \bibinfo  [0]{\@secondoftwo}%
\providecommand \bibfield  [0]{\@secondoftwo}%
\providecommand \translation [1]{[#1]}%
\providecommand \BibitemOpen [0]{}%
\providecommand \bibitemStop [0]{}%
\providecommand \bibitemNoStop [0]{.\EOS\space}%
\providecommand \EOS [0]{\spacefactor3000\relax}%
\providecommand \BibitemShut  [1]{\csname bibitem#1\endcsname}%
\let\auto@bib@innerbib\@empty
\bibitem [{\citenamefont {Maldacena}(1999)}]{Maldacena:1997re}%
  \BibitemOpen
  \bibfield  {author} {\bibinfo {author} {\bibfnamefont {J.~M.}\ \bibnamefont
  {Maldacena}},\ }\href {\doibase 10.1023/A:1026654312961} {\bibfield
  {journal} {\bibinfo  {journal} {Int. J. Theor. Phys.}\ }\textbf {\bibinfo
  {volume} {38}},\ \bibinfo {pages} {1113} (\bibinfo {year} {1999})},\ \Eprint
  {http://arxiv.org/abs/hep-th/9711200} {arXiv:hep-th/9711200} \BibitemShut
  {NoStop}%
\bibitem [{\citenamefont {Witten}(1998)}]{Witten:1998qj}%
  \BibitemOpen
  \bibfield  {author} {\bibinfo {author} {\bibfnamefont {E.}~\bibnamefont
  {Witten}},\ }\href {\doibase 10.4310/ATMP.1998.v2.n2.a2} {\bibfield
  {journal} {\bibinfo  {journal} {Adv. Theor. Math. Phys.}\ }\textbf {\bibinfo
  {volume} {2}},\ \bibinfo {pages} {253} (\bibinfo {year} {1998})},\ \Eprint
  {http://arxiv.org/abs/hep-th/9802150} {arXiv:hep-th/9802150} \BibitemShut
  {NoStop}%
\bibitem [{\citenamefont {Gubser}\ \emph {et~al.}(1998)\citenamefont {Gubser},
  \citenamefont {Klebanov},\ and\ \citenamefont {Polyakov}}]{Gubser:1998bc}%
  \BibitemOpen
  \bibfield  {author} {\bibinfo {author} {\bibfnamefont {S.}~\bibnamefont
  {Gubser}}, \bibinfo {author} {\bibfnamefont {I.~R.}\ \bibnamefont
  {Klebanov}}, \ and\ \bibinfo {author} {\bibfnamefont {A.~M.}\ \bibnamefont
  {Polyakov}},\ }\href {\doibase 10.1016/S0370-2693(98)00377-3} {\bibfield
  {journal} {\bibinfo  {journal} {Phys. Lett. B}\ }\textbf {\bibinfo {volume}
  {428}},\ \bibinfo {pages} {105} (\bibinfo {year} {1998})},\ \Eprint
  {http://arxiv.org/abs/hep-th/9802109} {arXiv:hep-th/9802109} \BibitemShut
  {NoStop}%
\bibitem [{\citenamefont {Karch}\ and\ \citenamefont
  {Katz}(2002)}]{Karch:2002sh}%
  \BibitemOpen
  \bibfield  {author} {\bibinfo {author} {\bibfnamefont {A.}~\bibnamefont
  {Karch}}\ and\ \bibinfo {author} {\bibfnamefont {E.}~\bibnamefont {Katz}},\
  }\href {\doibase 10.1088/1126-6708/2002/06/043} {\bibfield  {journal}
  {\bibinfo  {journal} {JHEP}\ }\textbf {\bibinfo {volume} {06}},\ \bibinfo
  {pages} {043} (\bibinfo {year} {2002})},\ \Eprint
  {http://arxiv.org/abs/hep-th/0205236} {arXiv:hep-th/0205236} \BibitemShut
  {NoStop}%
\bibitem [{\citenamefont {Kruczenski}\ \emph {et~al.}(2003)\citenamefont
  {Kruczenski}, \citenamefont {Mateos}, \citenamefont {Myers},\ and\
  \citenamefont {Winters}}]{Kruczenski:2003be}%
  \BibitemOpen
  \bibfield  {author} {\bibinfo {author} {\bibfnamefont {M.}~\bibnamefont
  {Kruczenski}}, \bibinfo {author} {\bibfnamefont {D.}~\bibnamefont {Mateos}},
  \bibinfo {author} {\bibfnamefont {R.~C.}\ \bibnamefont {Myers}}, \ and\
  \bibinfo {author} {\bibfnamefont {D.~J.}\ \bibnamefont {Winters}},\ }\href
  {\doibase 10.1088/1126-6708/2003/07/049} {\bibfield  {journal} {\bibinfo
  {journal} {JHEP}\ }\textbf {\bibinfo {volume} {07}},\ \bibinfo {pages} {049}
  (\bibinfo {year} {2003})},\ \Eprint {http://arxiv.org/abs/hep-th/0304032}
  {arXiv:hep-th/0304032} \BibitemShut {NoStop}%
\bibitem [{\citenamefont {Erdmenger}\ \emph {et~al.}(2008)\citenamefont
  {Erdmenger}, \citenamefont {Evans}, \citenamefont {Kirsch},\ and\
  \citenamefont {Threlfall}}]{Erdmenger:2007cm}%
  \BibitemOpen
  \bibfield  {author} {\bibinfo {author} {\bibfnamefont {J.}~\bibnamefont
  {Erdmenger}}, \bibinfo {author} {\bibfnamefont {N.}~\bibnamefont {Evans}},
  \bibinfo {author} {\bibfnamefont {I.}~\bibnamefont {Kirsch}}, \ and\ \bibinfo
  {author} {\bibfnamefont {E.}~\bibnamefont {Threlfall}},\ }\href {\doibase
  10.1140/epja/i2007-10540-1} {\bibfield  {journal} {\bibinfo  {journal} {Eur.
  Phys. J. A}\ }\textbf {\bibinfo {volume} {35}},\ \bibinfo {pages} {81}
  (\bibinfo {year} {2008})},\ \Eprint {http://arxiv.org/abs/0711.4467}
  {arXiv:0711.4467 [hep-th]} \BibitemShut {NoStop}%
\bibitem [{\citenamefont {Babington}\ \emph {et~al.}(2004)\citenamefont
  {Babington}, \citenamefont {Erdmenger}, \citenamefont {Evans}, \citenamefont
  {Guralnik},\ and\ \citenamefont {Kirsch}}]{Babington:2003vm}%
  \BibitemOpen
  \bibfield  {author} {\bibinfo {author} {\bibfnamefont {J.}~\bibnamefont
  {Babington}}, \bibinfo {author} {\bibfnamefont {J.}~\bibnamefont
  {Erdmenger}}, \bibinfo {author} {\bibfnamefont {N.~J.}\ \bibnamefont
  {Evans}}, \bibinfo {author} {\bibfnamefont {Z.}~\bibnamefont {Guralnik}}, \
  and\ \bibinfo {author} {\bibfnamefont {I.}~\bibnamefont {Kirsch}},\ }\href
  {\doibase 10.1103/PhysRevD.69.066007} {\bibfield  {journal} {\bibinfo
  {journal} {Phys. Rev. D}\ }\textbf {\bibinfo {volume} {69}},\ \bibinfo
  {pages} {066007} (\bibinfo {year} {2004})},\ \Eprint
  {http://arxiv.org/abs/hep-th/0306018} {arXiv:hep-th/0306018} \BibitemShut
  {NoStop}%
\bibitem [{\citenamefont {Kruczenski}\ \emph {et~al.}(2004)\citenamefont
  {Kruczenski}, \citenamefont {Mateos}, \citenamefont {Myers},\ and\
  \citenamefont {Winters}}]{Kruczenski:2003uq}%
  \BibitemOpen
  \bibfield  {author} {\bibinfo {author} {\bibfnamefont {M.}~\bibnamefont
  {Kruczenski}}, \bibinfo {author} {\bibfnamefont {D.}~\bibnamefont {Mateos}},
  \bibinfo {author} {\bibfnamefont {R.~C.}\ \bibnamefont {Myers}}, \ and\
  \bibinfo {author} {\bibfnamefont {D.~J.}\ \bibnamefont {Winters}},\ }\href
  {\doibase 10.1088/1126-6708/2004/05/041} {\bibfield  {journal} {\bibinfo
  {journal} {JHEP}\ }\textbf {\bibinfo {volume} {05}},\ \bibinfo {pages} {041}
  (\bibinfo {year} {2004})},\ \Eprint {http://arxiv.org/abs/hep-th/0311270}
  {arXiv:hep-th/0311270} \BibitemShut {NoStop}%
\bibitem [{\citenamefont {Cacciapaglia}\ \emph {et~al.}(2020)\citenamefont
  {Cacciapaglia}, \citenamefont {Pica},\ and\ \citenamefont
  {Sannino}}]{Cacciapaglia:2020kgq}%
  \BibitemOpen
  \bibfield  {author} {\bibinfo {author} {\bibfnamefont {G.}~\bibnamefont
  {Cacciapaglia}}, \bibinfo {author} {\bibfnamefont {C.}~\bibnamefont {Pica}},
  \ and\ \bibinfo {author} {\bibfnamefont {F.}~\bibnamefont {Sannino}},\
  }\href@noop {} {\bibfield  {journal} {\bibinfo  {journal} {Phys. Rept.}\
  }\textbf {\bibinfo {volume} {877}},\ \bibinfo {pages} {1} (\bibinfo {year}
  {2020})},\ \Eprint {http://arxiv.org/abs/2002.04914} {arXiv:2002.04914
  [hep-ph]} \BibitemShut {NoStop}%
\bibitem [{\citenamefont {Panico}\ and\ \citenamefont
  {Wulzer}(2016)}]{Panico:2015jxa}%
  \BibitemOpen
  \bibfield  {author} {\bibinfo {author} {\bibfnamefont {G.}~\bibnamefont
  {Panico}}\ and\ \bibinfo {author} {\bibfnamefont {A.}~\bibnamefont
  {Wulzer}},\ }\href {\doibase 10.1007/978-3-319-22617-0} {\emph {\bibinfo
  {title} {{The Composite Nambu-Goldstone Higgs}}}},\ Vol.\ \bibinfo {volume}
  {913}\ (\bibinfo  {publisher} {Springer},\ \bibinfo {year} {2016})\ \Eprint
  {http://arxiv.org/abs/1506.01961} {arXiv:1506.01961 [hep-ph]} \BibitemShut
  {NoStop}%
\bibitem [{\citenamefont {Arkani-Hamed}\ \emph {et~al.}(2002)\citenamefont
  {Arkani-Hamed}, \citenamefont {Cohen}, \citenamefont {Katz},\ and\
  \citenamefont {Nelson}}]{ArkaniHamed:2002qy}%
  \BibitemOpen
  \bibfield  {author} {\bibinfo {author} {\bibfnamefont {N.}~\bibnamefont
  {Arkani-Hamed}}, \bibinfo {author} {\bibfnamefont {A.}~\bibnamefont {Cohen}},
  \bibinfo {author} {\bibfnamefont {E.}~\bibnamefont {Katz}}, \ and\ \bibinfo
  {author} {\bibfnamefont {A.}~\bibnamefont {Nelson}},\ }\href {\doibase
  10.1088/1126-6708/2002/07/034} {\bibfield  {journal} {\bibinfo  {journal}
  {JHEP}\ }\textbf {\bibinfo {volume} {07}},\ \bibinfo {pages} {034} (\bibinfo
  {year} {2002})},\ \Eprint {http://arxiv.org/abs/hep-ph/0206021}
  {arXiv:hep-ph/0206021} \BibitemShut {NoStop}%
\bibitem [{\citenamefont {Bennett}\ \emph
  {et~al.}(2019{\natexlab{a}})\citenamefont {Bennett}, \citenamefont {Hong},
  \citenamefont {Lee}, \citenamefont {Lin}, \citenamefont {Lucini},
  \citenamefont {Mesiti}, \citenamefont {Piai}, \citenamefont {Rantaharju},\
  and\ \citenamefont {Vadacchino}}]{Bennett:2019cxd}%
  \BibitemOpen
  \bibfield  {author} {\bibinfo {author} {\bibfnamefont {E.}~\bibnamefont
  {Bennett}}, \bibinfo {author} {\bibfnamefont {D.~K.}\ \bibnamefont {Hong}},
  \bibinfo {author} {\bibfnamefont {J.-W.}\ \bibnamefont {Lee}}, \bibinfo
  {author} {\bibfnamefont {C.-J.~D.}\ \bibnamefont {Lin}}, \bibinfo {author}
  {\bibfnamefont {B.}~\bibnamefont {Lucini}}, \bibinfo {author} {\bibfnamefont
  {M.}~\bibnamefont {Mesiti}}, \bibinfo {author} {\bibfnamefont
  {M.}~\bibnamefont {Piai}}, \bibinfo {author} {\bibfnamefont {J.}~\bibnamefont
  {Rantaharju}}, \ and\ \bibinfo {author} {\bibfnamefont {D.}~\bibnamefont
  {Vadacchino}},\ }\href@noop {} {\  (\bibinfo {year} {2019}{\natexlab{a}})},\
  \Eprint {http://arxiv.org/abs/1912.06505} {arXiv:1912.06505 [hep-lat]}
  \BibitemShut {NoStop}%
\bibitem [{\citenamefont {Bennett}\ \emph
  {et~al.}(2019{\natexlab{b}})\citenamefont {Bennett}, \citenamefont {Hong},
  \citenamefont {Lee}, \citenamefont {Lin}, \citenamefont {Lucini},
  \citenamefont {Piai},\ and\ \citenamefont {Vadacchino}}]{Bennett:2019jzz}%
  \BibitemOpen
  \bibfield  {author} {\bibinfo {author} {\bibfnamefont {E.}~\bibnamefont
  {Bennett}}, \bibinfo {author} {\bibfnamefont {D.~K.}\ \bibnamefont {Hong}},
  \bibinfo {author} {\bibfnamefont {J.-W.}\ \bibnamefont {Lee}}, \bibinfo
  {author} {\bibfnamefont {C.-J.~D.}\ \bibnamefont {Lin}}, \bibinfo {author}
  {\bibfnamefont {B.}~\bibnamefont {Lucini}}, \bibinfo {author} {\bibfnamefont
  {M.}~\bibnamefont {Piai}}, \ and\ \bibinfo {author} {\bibfnamefont
  {D.}~\bibnamefont {Vadacchino}},\ }\href {\doibase 10.1007/JHEP12(2019)053}
  {\bibfield  {journal} {\bibinfo  {journal} {JHEP}\ }\textbf {\bibinfo
  {volume} {12}},\ \bibinfo {pages} {053} (\bibinfo {year}
  {2019}{\natexlab{b}})},\ \Eprint {http://arxiv.org/abs/1909.12662}
  {arXiv:1909.12662 [hep-lat]} \BibitemShut {NoStop}%
\bibitem [{\citenamefont {Ayyar}\ \emph
  {et~al.}(2018{\natexlab{a}})\citenamefont {Ayyar}, \citenamefont {Degrand},
  \citenamefont {Hackett}, \citenamefont {Jay}, \citenamefont {Neil},
  \citenamefont {Shamir},\ and\ \citenamefont {Svetitsky}}]{Ayyar:2018zuk}%
  \BibitemOpen
  \bibfield  {author} {\bibinfo {author} {\bibfnamefont {V.}~\bibnamefont
  {Ayyar}}, \bibinfo {author} {\bibfnamefont {T.}~\bibnamefont {Degrand}},
  \bibinfo {author} {\bibfnamefont {D.~C.}\ \bibnamefont {Hackett}}, \bibinfo
  {author} {\bibfnamefont {W.~I.}\ \bibnamefont {Jay}}, \bibinfo {author}
  {\bibfnamefont {E.~T.}\ \bibnamefont {Neil}}, \bibinfo {author}
  {\bibfnamefont {Y.}~\bibnamefont {Shamir}}, \ and\ \bibinfo {author}
  {\bibfnamefont {B.}~\bibnamefont {Svetitsky}},\ }\href {\doibase
  10.1103/PhysRevD.97.114505} {\bibfield  {journal} {\bibinfo  {journal} {Phys.
  Rev. D}\ }\textbf {\bibinfo {volume} {97}},\ \bibinfo {pages} {114505}
  (\bibinfo {year} {2018}{\natexlab{a}})},\ \Eprint
  {http://arxiv.org/abs/1801.05809} {arXiv:1801.05809 [hep-ph]} \BibitemShut
  {NoStop}%
\bibitem [{\citenamefont {Ayyar}\ \emph
  {et~al.}(2018{\natexlab{b}})\citenamefont {Ayyar}, \citenamefont {DeGrand},
  \citenamefont {Golterman}, \citenamefont {Hackett}, \citenamefont {Jay},
  \citenamefont {Neil}, \citenamefont {Shamir},\ and\ \citenamefont
  {Svetitsky}}]{Ayyar:2017qdf}%
  \BibitemOpen
  \bibfield  {author} {\bibinfo {author} {\bibfnamefont {V.}~\bibnamefont
  {Ayyar}}, \bibinfo {author} {\bibfnamefont {T.}~\bibnamefont {DeGrand}},
  \bibinfo {author} {\bibfnamefont {M.}~\bibnamefont {Golterman}}, \bibinfo
  {author} {\bibfnamefont {D.~C.}\ \bibnamefont {Hackett}}, \bibinfo {author}
  {\bibfnamefont {W.~I.}\ \bibnamefont {Jay}}, \bibinfo {author} {\bibfnamefont
  {E.~T.}\ \bibnamefont {Neil}}, \bibinfo {author} {\bibfnamefont
  {Y.}~\bibnamefont {Shamir}}, \ and\ \bibinfo {author} {\bibfnamefont
  {B.}~\bibnamefont {Svetitsky}},\ }\href {\doibase 10.1103/PhysRevD.97.074505}
  {\bibfield  {journal} {\bibinfo  {journal} {Phys. Rev. D}\ }\textbf {\bibinfo
  {volume} {97}},\ \bibinfo {pages} {074505} (\bibinfo {year}
  {2018}{\natexlab{b}})},\ \Eprint {http://arxiv.org/abs/1710.00806}
  {arXiv:1710.00806 [hep-lat]} \BibitemShut {NoStop}%
\bibitem [{\citenamefont {Ayyar}\ \emph {et~al.}(2019)\citenamefont {Ayyar},
  \citenamefont {DeGrand}, \citenamefont {Hackett}, \citenamefont {Jay},
  \citenamefont {Neil}, \citenamefont {Shamir},\ and\ \citenamefont
  {Svetitsky}}]{Ayyar:2018glg}%
  \BibitemOpen
  \bibfield  {author} {\bibinfo {author} {\bibfnamefont {V.}~\bibnamefont
  {Ayyar}}, \bibinfo {author} {\bibfnamefont {T.}~\bibnamefont {DeGrand}},
  \bibinfo {author} {\bibfnamefont {D.~C.}\ \bibnamefont {Hackett}}, \bibinfo
  {author} {\bibfnamefont {W.~I.}\ \bibnamefont {Jay}}, \bibinfo {author}
  {\bibfnamefont {E.~T.}\ \bibnamefont {Neil}}, \bibinfo {author}
  {\bibfnamefont {Y.}~\bibnamefont {Shamir}}, \ and\ \bibinfo {author}
  {\bibfnamefont {B.}~\bibnamefont {Svetitsky}},\ }\href {\doibase
  10.1103/PhysRevD.99.094502} {\bibfield  {journal} {\bibinfo  {journal} {Phys.
  Rev. D}\ }\textbf {\bibinfo {volume} {99}},\ \bibinfo {pages} {094502}
  (\bibinfo {year} {2019})},\ \Eprint {http://arxiv.org/abs/1812.02727}
  {arXiv:1812.02727 [hep-ph]} \BibitemShut {NoStop}%
\bibitem [{\citenamefont {Randall}\ and\ \citenamefont
  {Sundrum}(1999)}]{Randall:1999ee}%
  \BibitemOpen
  \bibfield  {author} {\bibinfo {author} {\bibfnamefont {L.}~\bibnamefont
  {Randall}}\ and\ \bibinfo {author} {\bibfnamefont {R.}~\bibnamefont
  {Sundrum}},\ }\href {\doibase 10.1103/PhysRevLett.83.3370} {\bibfield
  {journal} {\bibinfo  {journal} {Phys. Rev. Lett.}\ }\textbf {\bibinfo
  {volume} {83}},\ \bibinfo {pages} {3370} (\bibinfo {year} {1999})},\ \Eprint
  {http://arxiv.org/abs/hep-ph/9905221} {arXiv:hep-ph/9905221} \BibitemShut
  {NoStop}%
\bibitem [{\citenamefont {Contino}\ \emph {et~al.}(2003)\citenamefont
  {Contino}, \citenamefont {Nomura},\ and\ \citenamefont
  {Pomarol}}]{Contino:2003ve}%
  \BibitemOpen
  \bibfield  {author} {\bibinfo {author} {\bibfnamefont {R.}~\bibnamefont
  {Contino}}, \bibinfo {author} {\bibfnamefont {Y.}~\bibnamefont {Nomura}}, \
  and\ \bibinfo {author} {\bibfnamefont {A.}~\bibnamefont {Pomarol}},\ }\href
  {\doibase 10.1016/j.nuclphysb.2003.08.027} {\bibfield  {journal} {\bibinfo
  {journal} {Nucl. Phys. B}\ }\textbf {\bibinfo {volume} {671}},\ \bibinfo
  {pages} {148} (\bibinfo {year} {2003})},\ \Eprint
  {http://arxiv.org/abs/hep-ph/0306259} {arXiv:hep-ph/0306259} \BibitemShut
  {NoStop}%
\bibitem [{\citenamefont {Agashe}\ \emph {et~al.}(2005)\citenamefont {Agashe},
  \citenamefont {Contino},\ and\ \citenamefont {Pomarol}}]{Agashe:2004rs}%
  \BibitemOpen
  \bibfield  {author} {\bibinfo {author} {\bibfnamefont {K.}~\bibnamefont
  {Agashe}}, \bibinfo {author} {\bibfnamefont {R.}~\bibnamefont {Contino}}, \
  and\ \bibinfo {author} {\bibfnamefont {A.}~\bibnamefont {Pomarol}},\ }\href
  {\doibase 10.1016/j.nuclphysb.2005.04.035} {\bibfield  {journal} {\bibinfo
  {journal} {Nucl. Phys. B}\ }\textbf {\bibinfo {volume} {719}},\ \bibinfo
  {pages} {165} (\bibinfo {year} {2005})},\ \Eprint
  {http://arxiv.org/abs/hep-ph/0412089} {arXiv:hep-ph/0412089} \BibitemShut
  {NoStop}%
\bibitem [{\citenamefont {Barnard}\ \emph {et~al.}(2014)\citenamefont
  {Barnard}, \citenamefont {Gherghetta},\ and\ \citenamefont
  {Ray}}]{Barnard:2013zea}%
  \BibitemOpen
  \bibfield  {author} {\bibinfo {author} {\bibfnamefont {J.}~\bibnamefont
  {Barnard}}, \bibinfo {author} {\bibfnamefont {T.}~\bibnamefont {Gherghetta}},
  \ and\ \bibinfo {author} {\bibfnamefont {T.~S.}\ \bibnamefont {Ray}},\ }\href
  {\doibase 10.1007/JHEP02(2014)002} {\bibfield  {journal} {\bibinfo  {journal}
  {JHEP}\ }\textbf {\bibinfo {volume} {02}},\ \bibinfo {pages} {002} (\bibinfo
  {year} {2014})},\ \Eprint {http://arxiv.org/abs/1311.6562} {arXiv:1311.6562
  [hep-ph]} \BibitemShut {NoStop}%
\bibitem [{\citenamefont {Ferretti}(2014)}]{Ferretti:2014qta}%
  \BibitemOpen
  \bibfield  {author} {\bibinfo {author} {\bibfnamefont {G.}~\bibnamefont
  {Ferretti}},\ }\href {\doibase 10.1007/JHEP06(2014)142} {\bibfield  {journal}
  {\bibinfo  {journal} {JHEP}\ }\textbf {\bibinfo {volume} {06}},\ \bibinfo
  {pages} {142} (\bibinfo {year} {2014})},\ \Eprint
  {http://arxiv.org/abs/1404.7137} {arXiv:1404.7137 [hep-ph]} \BibitemShut
  {NoStop}%
\bibitem [{\citenamefont {Ferretti}(2016)}]{Ferretti:2016upr}%
  \BibitemOpen
  \bibfield  {author} {\bibinfo {author} {\bibfnamefont {G.}~\bibnamefont
  {Ferretti}},\ }\href {\doibase 10.1007/JHEP06(2016)107} {\bibfield  {journal}
  {\bibinfo  {journal} {JHEP}\ }\textbf {\bibinfo {volume} {06}},\ \bibinfo
  {pages} {107} (\bibinfo {year} {2016})},\ \Eprint
  {http://arxiv.org/abs/1604.06467} {arXiv:1604.06467 [hep-ph]} \BibitemShut
  {NoStop}%
\bibitem [{\citenamefont {Evans}\ and\ \citenamefont
  {Tuominen}(2013)}]{Evans:2013vca}%
  \BibitemOpen
  \bibfield  {author} {\bibinfo {author} {\bibfnamefont {N.}~\bibnamefont
  {Evans}}\ and\ \bibinfo {author} {\bibfnamefont {K.}~\bibnamefont
  {Tuominen}},\ }\href {\doibase 10.1103/PhysRevD.87.086003} {\bibfield
  {journal} {\bibinfo  {journal} {Phys. Rev. D}\ }\textbf {\bibinfo {volume}
  {87}},\ \bibinfo {pages} {086003} (\bibinfo {year} {2013})},\ \Eprint
  {http://arxiv.org/abs/1302.4553} {arXiv:1302.4553 [hep-ph]} \BibitemShut
  {NoStop}%
\bibitem [{\citenamefont {Kaplan}(1991)}]{Kaplan:1991dc}%
  \BibitemOpen
  \bibfield  {author} {\bibinfo {author} {\bibfnamefont {D.~B.}\ \bibnamefont
  {Kaplan}},\ }\href {\doibase 10.1016/S0550-3213(05)80021-5} {\bibfield
  {journal} {\bibinfo  {journal} {Nucl. Phys. B}\ }\textbf {\bibinfo {volume}
  {365}},\ \bibinfo {pages} {259} (\bibinfo {year} {1991})}\BibitemShut
  {NoStop}%
\bibitem [{\citenamefont {Aaboud}\ \emph {et~al.}(2018)\citenamefont {Aaboud}
  \emph {et~al.}}]{Aaboud:2018pii}%
  \BibitemOpen
  \bibfield  {author} {\bibinfo {author} {\bibfnamefont {M.}~\bibnamefont
  {Aaboud}} \emph {et~al.} (\bibinfo {collaboration} {ATLAS}),\ }\href
  {\doibase 10.1103/PhysRevLett.121.211801} {\bibfield  {journal} {\bibinfo
  {journal} {Phys. Rev. Lett.}\ }\textbf {\bibinfo {volume} {121}},\ \bibinfo
  {pages} {211801} (\bibinfo {year} {2018})},\ \Eprint
  {http://arxiv.org/abs/1808.02343} {arXiv:1808.02343 [hep-ex]} \BibitemShut
  {NoStop}%
\bibitem [{\citenamefont {Sirunyan}\ \emph {et~al.}(2019)\citenamefont
  {Sirunyan} \emph {et~al.}}]{Sirunyan:2019sza}%
  \BibitemOpen
  \bibfield  {author} {\bibinfo {author} {\bibfnamefont {A.~M.}\ \bibnamefont
  {Sirunyan}} \emph {et~al.} (\bibinfo {collaboration} {CMS}),\ }\href
  {\doibase 10.1103/PhysRevD.100.072001} {\bibfield  {journal} {\bibinfo
  {journal} {Phys. Rev. D}\ }\textbf {\bibinfo {volume} {100}},\ \bibinfo
  {pages} {072001} (\bibinfo {year} {2019})},\ \Eprint
  {http://arxiv.org/abs/1906.11903} {arXiv:1906.11903 [hep-ex]} \BibitemShut
  {NoStop}%
\bibitem [{Note1()}]{Note1}%
  \BibitemOpen
  \bibinfo {note} {However, these bounds can be lower if non-standard decays
  are important \cite {Cacciapaglia:2019zmj}.}\BibitemShut {Stop}%
\bibitem [{\citenamefont {Abt}\ \emph {et~al.}(2019)\citenamefont {Abt},
  \citenamefont {Erdmenger}, \citenamefont {Evans},\ and\ \citenamefont
  {Rigatos}}]{Abt:2019tas}%
  \BibitemOpen
  \bibfield  {author} {\bibinfo {author} {\bibfnamefont {R.}~\bibnamefont
  {Abt}}, \bibinfo {author} {\bibfnamefont {J.}~\bibnamefont {Erdmenger}},
  \bibinfo {author} {\bibfnamefont {N.}~\bibnamefont {Evans}}, \ and\ \bibinfo
  {author} {\bibfnamefont {K.~S.}\ \bibnamefont {Rigatos}},\ }\href {\doibase
  10.1007/JHEP11(2019)160} {\bibfield  {journal} {\bibinfo  {journal} {JHEP}\
  }\textbf {\bibinfo {volume} {11}},\ \bibinfo {pages} {160} (\bibinfo {year}
  {2019})},\ \Eprint {http://arxiv.org/abs/1907.09489} {arXiv:1907.09489
  [hep-th]} \BibitemShut {NoStop}%
\bibitem [{\citenamefont {Witten}(2001)}]{Witten:2001ua}%
  \BibitemOpen
  \bibfield  {author} {\bibinfo {author} {\bibfnamefont {E.}~\bibnamefont
  {Witten}},\ }\href@noop {} {\  (\bibinfo {year} {2001})},\ \Eprint
  {http://arxiv.org/abs/hep-th/0112258} {arXiv:hep-th/0112258} \BibitemShut
  {NoStop}%
\bibitem [{\citenamefont {Alho}\ \emph {et~al.}(2013)\citenamefont {Alho},
  \citenamefont {Evans},\ and\ \citenamefont {Tuominen}}]{Alho:2013dka}%
  \BibitemOpen
  \bibfield  {author} {\bibinfo {author} {\bibfnamefont {T.}~\bibnamefont
  {Alho}}, \bibinfo {author} {\bibfnamefont {N.}~\bibnamefont {Evans}}, \ and\
  \bibinfo {author} {\bibfnamefont {K.}~\bibnamefont {Tuominen}},\ }\href
  {\doibase 10.1103/PhysRevD.88.105016} {\bibfield  {journal} {\bibinfo
  {journal} {Phys. Rev. D}\ }\textbf {\bibinfo {volume} {88}},\ \bibinfo
  {pages} {105016} (\bibinfo {year} {2013})},\ \Eprint
  {http://arxiv.org/abs/1307.4896} {arXiv:1307.4896 [hep-ph]} \BibitemShut
  {NoStop}%
\bibitem [{\citenamefont {Alvares}\ \emph {et~al.}(2012)\citenamefont
  {Alvares}, \citenamefont {Evans},\ and\ \citenamefont
  {Kim}}]{Alvares:2012kr}%
  \BibitemOpen
  \bibfield  {author} {\bibinfo {author} {\bibfnamefont {R.}~\bibnamefont
  {Alvares}}, \bibinfo {author} {\bibfnamefont {N.}~\bibnamefont {Evans}}, \
  and\ \bibinfo {author} {\bibfnamefont {K.-Y.}\ \bibnamefont {Kim}},\ }\href
  {\doibase 10.1103/PhysRevD.86.026008} {\bibfield  {journal} {\bibinfo
  {journal} {Phys. Rev. D}\ }\textbf {\bibinfo {volume} {86}},\ \bibinfo
  {pages} {026008} (\bibinfo {year} {2012})},\ \Eprint
  {http://arxiv.org/abs/1204.2474} {arXiv:1204.2474 [hep-ph]} \BibitemShut
  {NoStop}%
\bibitem [{\citenamefont {Erdmenger}\ \emph {et~al.}(2015)\citenamefont
  {Erdmenger}, \citenamefont {Evans},\ and\ \citenamefont
  {Scott}}]{Erdmenger:2014fxa}%
  \BibitemOpen
  \bibfield  {author} {\bibinfo {author} {\bibfnamefont {J.}~\bibnamefont
  {Erdmenger}}, \bibinfo {author} {\bibfnamefont {N.}~\bibnamefont {Evans}}, \
  and\ \bibinfo {author} {\bibfnamefont {M.}~\bibnamefont {Scott}},\ }\href
  {\doibase 10.1103/PhysRevD.91.085004} {\bibfield  {journal} {\bibinfo
  {journal} {Phys. Rev. D}\ }\textbf {\bibinfo {volume} {91}},\ \bibinfo
  {pages} {085004} (\bibinfo {year} {2015})},\ \Eprint
  {http://arxiv.org/abs/1412.3165} {arXiv:1412.3165 [hep-ph]} \BibitemShut
  {NoStop}%
\bibitem [{\citenamefont {Erlich}\ \emph {et~al.}(2005)\citenamefont {Erlich},
  \citenamefont {Katz}, \citenamefont {Son},\ and\ \citenamefont
  {Stephanov}}]{Erlich:2005qh}%
  \BibitemOpen
  \bibfield  {author} {\bibinfo {author} {\bibfnamefont {J.}~\bibnamefont
  {Erlich}}, \bibinfo {author} {\bibfnamefont {E.}~\bibnamefont {Katz}},
  \bibinfo {author} {\bibfnamefont {D.~T.}\ \bibnamefont {Son}}, \ and\
  \bibinfo {author} {\bibfnamefont {M.~A.}\ \bibnamefont {Stephanov}},\ }\href
  {\doibase 10.1103/PhysRevLett.95.261602} {\bibfield  {journal} {\bibinfo
  {journal} {Phys. Rev. Lett.}\ }\textbf {\bibinfo {volume} {95}},\ \bibinfo
  {pages} {261602} (\bibinfo {year} {2005})},\ \Eprint
  {http://arxiv.org/abs/hep-ph/0501128} {arXiv:hep-ph/0501128} \BibitemShut
  {NoStop}%
\bibitem [{\citenamefont {Da~Rold}\ and\ \citenamefont
  {Pomarol}(2005)}]{DaRold:2005mxj}%
  \BibitemOpen
  \bibfield  {author} {\bibinfo {author} {\bibfnamefont {L.}~\bibnamefont
  {Da~Rold}}\ and\ \bibinfo {author} {\bibfnamefont {A.}~\bibnamefont
  {Pomarol}},\ }\href {\doibase 10.1016/j.nuclphysb.2005.05.009} {\bibfield
  {journal} {\bibinfo  {journal} {Nucl. Phys. B}\ }\textbf {\bibinfo {volume}
  {721}},\ \bibinfo {pages} {79} (\bibinfo {year} {2005})},\ \Eprint
  {http://arxiv.org/abs/hep-ph/0501218} {arXiv:hep-ph/0501218} \BibitemShut
  {NoStop}%
\bibitem [{Note2()}]{Note2}%
  \BibitemOpen
  \bibinfo {note} {${\mathchoice {{\setbox \z@ \hbox {$\mathsurround \z@
  \displaystyle D$}\setbox \tw@ \hbox {$\mathsurround \z@ \displaystyle
  /$}\dimen 4\wd \z@ \dimen@ \ht \tw@ \advance \dimen@ -\dp \tw@ \advance
  \dimen@ -\ht \z@ \advance \dimen@ \dp \z@ \divide \dimen@ \tw@ \advance
  \dimen@ -0\ht \tw@ \advance \dimen@ -0\dp \tw@ \dimen@ii .08\wd \z@ \raise
  -\dimen@ \hbox to\dimen 4{\hss \kern \dimen@ii \box \tw@ \kern -\dimen@ii
  \hss }\hbox to\z@ {\hss \hbox to\dimen 4{\hss \box \z@ \hss }}}}{{\setbox \z@
  \hbox {$\mathsurround \z@ \textstyle D$}\setbox \tw@ \hbox {$\mathsurround
  \z@ \textstyle /$}\dimen 4\wd \z@ \dimen@ \ht \tw@ \advance \dimen@ -\dp \tw@
  \advance \dimen@ -\ht \z@ \advance \dimen@ \dp \z@ \divide \dimen@ \tw@
  \advance \dimen@ -0\ht \tw@ \advance \dimen@ -0\dp \tw@ \dimen@ii .08\wd \z@
  \raise -\dimen@ \hbox to\dimen 4{\hss \kern \dimen@ii \box \tw@ \kern
  -\dimen@ii \hss }\hbox to\z@ {\hss \hbox to\dimen 4{\hss \box \z@ \hss
  }}}}{{\setbox \z@ \hbox {$\mathsurround \z@ \scriptstyle D$}\setbox \tw@
  \hbox {$\mathsurround \z@ \scriptstyle /$}\dimen 4\wd \z@ \dimen@ \ht \tw@
  \advance \dimen@ -\dp \tw@ \advance \dimen@ -\ht \z@ \advance \dimen@ \dp \z@
  \divide \dimen@ \tw@ \advance \dimen@ -0\ht \tw@ \advance \dimen@ -0\dp \tw@
  \dimen@ii .08\wd \z@ \raise -\dimen@ \hbox to\dimen 4{\hss \kern \dimen@ii
  \box \tw@ \kern -\dimen@ii \hss }\hbox to\z@ {\hss \hbox to\dimen 4{\hss \box
  \z@ \hss }}}}{{\setbox \z@ \hbox {$\mathsurround \z@ \scriptscriptstyle
  D$}\setbox \tw@ \hbox {$\mathsurround \z@ \scriptscriptstyle /$}\dimen 4\wd
  \z@ \dimen@ \ht \tw@ \advance \dimen@ -\dp \tw@ \advance \dimen@ -\ht \z@
  \advance \dimen@ \dp \z@ \divide \dimen@ \tw@ \advance \dimen@ -0\ht \tw@
  \advance \dimen@ -0\dp \tw@ \dimen@ii .08\wd \z@ \raise -\dimen@ \hbox
  to\dimen 4{\hss \kern \dimen@ii \box \tw@ \kern -\dimen@ii \hss }\hbox to\z@
  {\hss \hbox to\dimen 4{\hss \box \z@ \hss }}}}}_{\protect \text {AAdS}}$ is
  the Dirac operator evaluated on \protect \cref {eq: bckgrnd}}\BibitemShut
  {NoStop}%
\bibitem [{\citenamefont {Breitenlohner}\ and\ \citenamefont
  {Freedman}(1982)}]{Breitenlohner:1982jf}%
  \BibitemOpen
  \bibfield  {author} {\bibinfo {author} {\bibfnamefont {P.}~\bibnamefont
  {Breitenlohner}}\ and\ \bibinfo {author} {\bibfnamefont {D.~Z.}\ \bibnamefont
  {Freedman}},\ }\href {\doibase 10.1016/0003-4916(82)90116-6} {\bibfield
  {journal} {\bibinfo  {journal} {Annals Phys.}\ }\textbf {\bibinfo {volume}
  {144}},\ \bibinfo {pages} {249} (\bibinfo {year} {1982})}\BibitemShut
  {NoStop}%
\bibitem [{Note3()}]{Note3}%
  \BibitemOpen
  \bibinfo {note} {For all the group theory factors we have used the
  Mathematica application LieART \cite {Feger:2012bs}}\BibitemShut {NoStop}%
\bibitem [{foo()}]{footnote}%
  \BibitemOpen
  \href@noop {} {}\bibinfo {note} {For a variable $C=(A/B)^{1/2}$ we translate
  errors as $dC = 1/2 (dA/(AB)^{1/2} + A^{1/2} dB/B^{3/2})$}\BibitemShut
  {NoStop}%
\bibitem [{\citenamefont {Lewis}\ \emph {et~al.}(2012)\citenamefont {Lewis},
  \citenamefont {Pica},\ and\ \citenamefont {Sannino}}]{Lewis:2011zb}%
  \BibitemOpen
  \bibfield  {author} {\bibinfo {author} {\bibfnamefont {R.}~\bibnamefont
  {Lewis}}, \bibinfo {author} {\bibfnamefont {C.}~\bibnamefont {Pica}}, \ and\
  \bibinfo {author} {\bibfnamefont {F.}~\bibnamefont {Sannino}},\ }\href
  {\doibase 10.1103/PhysRevD.85.014504} {\bibfield  {journal} {\bibinfo
  {journal} {Phys. Rev. D}\ }\textbf {\bibinfo {volume} {85}},\ \bibinfo
  {pages} {014504} (\bibinfo {year} {2012})},\ \Eprint
  {http://arxiv.org/abs/1109.3513} {arXiv:1109.3513 [hep-ph]} \BibitemShut
  {NoStop}%
\bibitem [{\citenamefont {Kirsch}(2006)}]{Kirsch:2006he}%
  \BibitemOpen
  \bibfield  {author} {\bibinfo {author} {\bibfnamefont {I.}~\bibnamefont
  {Kirsch}},\ }\href {\doibase 10.1088/1126-6708/2006/09/052} {\bibfield
  {journal} {\bibinfo  {journal} {JHEP}\ }\textbf {\bibinfo {volume} {09}},\
  \bibinfo {pages} {052} (\bibinfo {year} {2006})},\ \Eprint
  {http://arxiv.org/abs/hep-th/0607205} {arXiv:hep-th/0607205} \BibitemShut
  {NoStop}%
\bibitem [{\citenamefont {Erdmenger}\ \emph {et~al.}(2020)\citenamefont
  {Erdmenger}, \citenamefont {Evans}, \citenamefont {Porod},\ and\
  \citenamefont {Rigatos}}]{Erdmenger:2020flu}%
  \BibitemOpen
  \bibfield  {author} {\bibinfo {author} {\bibfnamefont {J.}~\bibnamefont
  {Erdmenger}}, \bibinfo {author} {\bibfnamefont {N.}~\bibnamefont {Evans}},
  \bibinfo {author} {\bibfnamefont {W.}~\bibnamefont {Porod}}, \ and\ \bibinfo
  {author} {\bibfnamefont {K.~S.}\ \bibnamefont {Rigatos}},\ }\href@noop {} {\
  (\bibinfo {year} {2020})},\ \Eprint {http://arxiv.org/abs/2010.10279}
  {arXiv:2010.10279 [hep-ph]} \BibitemShut {NoStop}%
\bibitem [{Note4()}]{Note4}%
  \BibitemOpen
  \bibinfo {note} {To compute the $Z_3$ factor, we would need a cubic term in
  \protect \textup {\hbox {\mathsurround \z@ \protect \normalfont
  (\ignorespaces \ref {eq: general_action}\unskip \@@italiccorr )}}. However,
  its coupling is not determined in the bottom-up approach. We use the estimate
  given in \protect \textup {\hbox {\mathsurround \z@ \protect \normalfont
  (\ignorespaces \ref {eq: top_mass_1a}\unskip \@@italiccorr )}}.}\BibitemShut
  {Stop}%
\bibitem [{\citenamefont {Buyukdag}(2019)}]{Buyukdag:2019ouy}%
  \BibitemOpen
  \bibfield  {author} {\bibinfo {author} {\bibfnamefont {Y.}~\bibnamefont
  {Buyukdag}},\ }\href@noop {} {\  (\bibinfo {year} {2019})},\ \Eprint
  {http://arxiv.org/abs/1911.12328} {arXiv:1911.12328 [hep-th]} \BibitemShut
  {NoStop}%
\bibitem [{\citenamefont {Cacciapaglia}\ \emph {et~al.}(2019)\citenamefont
  {Cacciapaglia}, \citenamefont {Flacke}, \citenamefont {Park},\ and\
  \citenamefont {Zhang}}]{Cacciapaglia:2019zmj}%
  \BibitemOpen
  \bibfield  {author} {\bibinfo {author} {\bibfnamefont {G.}~\bibnamefont
  {Cacciapaglia}}, \bibinfo {author} {\bibfnamefont {T.}~\bibnamefont
  {Flacke}}, \bibinfo {author} {\bibfnamefont {M.}~\bibnamefont {Park}}, \ and\
  \bibinfo {author} {\bibfnamefont {M.}~\bibnamefont {Zhang}},\ }\href
  {\doibase 10.1016/j.physletb.2019.135015} {\bibfield  {journal} {\bibinfo
  {journal} {Phys. Lett. B}\ }\textbf {\bibinfo {volume} {798}},\ \bibinfo
  {pages} {135015} (\bibinfo {year} {2019})},\ \Eprint
  {http://arxiv.org/abs/1908.07524} {arXiv:1908.07524 [hep-ph]} \BibitemShut
  {NoStop}%
\bibitem [{\citenamefont {Feger}\ and\ \citenamefont
  {Kephart}(2015)}]{Feger:2012bs}%
  \BibitemOpen
  \bibfield  {author} {\bibinfo {author} {\bibfnamefont {R.}~\bibnamefont
  {Feger}}\ and\ \bibinfo {author} {\bibfnamefont {T.~W.}\ \bibnamefont
  {Kephart}},\ }\href {\doibase 10.1016/j.cpc.2014.12.023} {\bibfield
  {journal} {\bibinfo  {journal} {Comput. Phys. Commun.}\ }\textbf {\bibinfo
  {volume} {192}},\ \bibinfo {pages} {166} (\bibinfo {year} {2015})},\ \Eprint
  {http://arxiv.org/abs/1206.6379} {arXiv:1206.6379 [math-ph]} \BibitemShut
  {NoStop}%
\end{thebibliography}%
\onecolumngrid
\end{document}